\documentstyle[12pt]{article}
\newsavebox{\prooffig}

\def\del        {  \partial  }
\def\half       {  {1\over 2}  }
\def\Tr      {{\rm Tr}}
\def\lsim    {\lower .65ex \hbox{\ $\stackrel{<}{\sim}$\ } }
\def\gsim    {\lower .65ex \hbox{\ $\stackrel{>}{\sim}$\ } }
\def\vecii#1#2      {  \left(\begin{array}{c}#1\\#2\end{array}\right)  }
\def\veciii#1#2#3   {  \left(\begin{array}{c}#1\\#2\\#3\end{array}
                     \right)  }
\def\veciv#1#2#3#4  {  \left(\begin{array}{c}#1\\#2\\#3\\#4
                                 \end{array}\right)  }
\def\vecfv#1#2#3#4#5 {  \left(\begin{array}{c}#1\\#2\\#3\\#4\\#5
                                 \end{array}\right)  }

\def\matrixii#1#2#3#4            {  \left(\begin{array}{cc}#1&#2\\#3&#4
                                       \end{array}\right) }
\def\matrixiii#1#2#3#4#5#6#7#8#9 {  \left(\begin{array}{ccc}#1&#2&#3\\
                                     #4&#5&#6\\#7&#8&#9\end{array}
                               \right)  }
\def\mativ#1#2#3#4               {  \left(\begin{array}{cccc}
                                       #1\\#2\\#3\\#4\end{array}\right) }
\def\matv#1#2#3#4#5              {  \left(\begin{array}{ccccc}
                                     #1\\#2\\#3\\#4\\#5\end{array}
                              \right)  }
\def\eqabegin         {  \begin{eqnarray}  }
\def\eqaend           {  \end{eqnarray}  }
\def\nn               {  \nonumber  }
\def\bracetwo#1#2     {  \left\{ \begin{array}{l} #1 \\ #2 \end{array}
                         \right.  }
\def\bracetwocases#1#2#3#4  {   \left\{ \begin{array}{ll} #1 &
                                 \qquad #2 \\
                                 #3 & \qquad #4 \end{array} \right.  }
\def\bracebegin#1     {  \left\{ \begin{array}{#1}   }
\def\braceend         {  \end{array}\right.   }
\def\parn              {  \par\noindent }

\def\parmedskip        {  \par\medskip  }

\def\parbigskipn        {  \par\bigskip\noindent  }
\def\parmedskipn        {  \par\medskip\noindent  }

\def\parag#1           {\paragraph{#1} \mbox{ }\parmedskip\noindent}
\def\boxit#1#2      {  \vbox{\hrule\hbox{ \hskip -4.1pt \vrule\kern3pt

                     \vbox
                    {  \hsize #1 \strut\kern3pt #2 \kern3pt\strut  }
                       \kern3pt  \vrule} \hrule  } }
\def\centerbox#1#2  {  \mbox{  }\par\bigskip  \hfil \boxit{#1}{#2} \hfil
                       \par\bigskip\noindent }
\def\rightbox#1#2   {  \hfill\boxit{#1}{#2}  }
\def\leftbox#1#2    {  \boxit{#1}{#2}  }
\def\fullbox#1      {  \boxit{\textwidth}{#1}  }
\def\rightfigspacebegin  {  \par\noindent\begin{minipage}[t]{10cm}  }
\def\rightfigspaceend    {  \end{minipage}\par\noindent  }
\def\leftfigspacebegin   {  \par\noindent
                             \hspace*{10cm}\begin{minipage}[t]{6cm} }
\def\leftfigspaceend     {  \end{minipage}\par\noindent  }
\def\titleandfile#1#2   {  \begin{center}{\Large\bf #1}\end{center}
                            \par\begin{flushright} #2 \end{flushright}  }
\def\msection#1      {  \begin{center} \section{#1} \end{center}   }
\def\nsection#1      {  \let\boldface\bf \def\bf{} \section{#1}
                           \let\bf\boldface   }
\def\mnsection#1     {  \begin{center} \nsection{#1} \end{center}  }
\def\capsection#1    {  \let\boldface\bf \def\bf{\sc} \section{#1}
                           \let\bf\boldface   }
\def\mcapsection#1   {  \begin{center} \capsection{#1} \end{center} }

\newcommand{\nullify}[1]{}
\def\papertitlepage{\baselineskip 3.5ex \thispagestyle{empty}}
\def\Title#1{\baselineskip 1cm \vspace{1.5cm}\begin{center}
 {\Large\bf #1} \end{center}
\vspace{0.5cm}}
\def\Authors#1{\begin{center} {\it #1} \end{center}}
\def\Abstract{\vspace{1.0cm}\begin{center} {\large\bf Abstract}
           \end{center} \par\bigskip}
\def\Number#1#2#3#4{\hfill \begin{minipage}{4.2cm} #1 \parn #2
              \parn #3
              \parn #4 \end{minipage}}
\renewcommand{\thefootnote}{\fnsymbol{footnote}}
\renewenvironment{thebibliography}{\pagebreak[3]\par\vspace{0.6em}
\begin{flushleft}{\large \bf References}\end{flushleft}
\vspace{-1.0em}

\begin{enumerate}\if@twocolumn\baselineskip=0.6em\itemsep -0.2em
\else\itemsep -0.2em\fi\labelsep 0.1em}{\end{enumerate}}
\begin{document}
\papertitlepage
\vspace*{-0.85cm}
\Number{OU-HET 274}{UT-Komaba/97-11}{hep-th/9708039}{August 1997}
\Title{{\sc Topological Matrix Model}}
\vspace{0.2cm}
\Authors{{\sc Shinji~Hirano
\footnote[2]{hirano@funpth.phys.sci.osaka-u.ac.jp}
 and
 \sc Mitsuhiro~Kato
\footnote[3]{kato@hep1.c.u-tokyo.ac.jp}
\\ }
\vskip 2.5ex
${}^{\dag}$ Department of Physics, Graduate School of Science, \\
 Osaka University, Toyonaka, Osaka, 560 Japan \\
  and \\
${}^{\ddag}$ Institute of Physics, University of Tokyo, \\ Komaba, Meguro-ku,
Tokyo 153 Japan \\ }
\baselineskip .7cm
\vspace*{-0.3cm}
\Abstract {Starting from the primal principle based on the noncommutative
nature of ($9+1$)-dimensional spacetime, we construct a topologically twisted
version of the supersymmetric reduced model with a certain modification. Our
formulation automatically provides extra $1+1$ dimensions, thereby the
dimensions of spacetime are promoted to $10+2$. With a suitable gauge choice,
we can reduce the model with ($10+2$)-dimensional spacetime to the one with
($9+1$)-dimensions and thus we regard this gauge as the light-cone gauge. It is
 suggested that the model so obtained would describe the light-cone F-theory.
{}From this viewpoint we argue the relation of the reduced model to the matrix
model of M-theory and the $SL(2,Z)$ symmetry of type IIB string theory. We also
discuss the general covariance of the matrix model in a broken phase, and make
some comments on the background independence.}
\newpage


\renewcommand{\thefootnote}{\arabic{footnote}}
\section{{\sc Introduction}}

The background independence is the most significant implication of the
inclusion
of quantum gravity in string theory. The geometry of spacetime should not be
set
up a priori, rather it is generated by a highly nonperturbative effect, the
condensation of strings. Thus the understanding of the background independence
is promisingly the key ingredient to seek the underlying principle of
nonperturbative string theory.

Although the matrix model of M-theory \cite{BFSS} and IIB matrix models
\cite{IKKT1}\nocite{NBI}\nocite{Shild}--\cite{CheTsey} may provide possible
descriptions of the underlying theory of
string theory, they are still lacking the fundamental principle and the
understanding of the background independence.  The matrix model of M-theory,
however, might have provided some clues to these problems. An indication is the
existence of the more fundamental degrees of freedom, D-particles or partons,
from which strings are constituted. As discussed in \cite{DKPS}, this suggests
the emergence of the scales shorter than string length. More remarkably they
exhibit the {\it noncommutative} nature of spacetime \cite{WittenBS}, and thus
they are considered to be inherently non-local or fuzzy objects. This property
is quite desirable to keep the fine upshots of perturbative string theory, such
as the ultraviolet finiteness and T-duality \cite{KYSS}, stemming from the
extended character of strings.

These observations tempted us to look for the underlying principle based on the
noncommutativity of spacetime. In the present work, we consider (anti-)
D-instantons as the fundamental degrees of freedom instead of
D-particles. D-particles should be constituted from D-instantons, just
in the same spirit as the strings from D-particles in the matrix model of
M-theory. The fuzzy instantons are represented by pure matrices, that is,
matrices without continuous parameters. They are nothing other than the
coordinates of a noncommutative spacetime. We shall take them as the only
elements to construct our model, and impose a symmetry of arbitrary
deformations
of matrices. Thus our construction of the model follows that of a topological
quantum field theory \cite{WittenTFT}.

In section \ref{sec:TM} we will construct a topologically twisted version of
the
supersymmetric reduced model with a certain modification. We argue that the
critical dimensions of spacetime, including the signature, would be restricted
to some extent by our formulation of the model. We make a brief remark on an
additional term which is missing in the supersymmetric reduced model.

Our formulation suggests that the model we constructed would provide a
possible description of F-theory \cite{Vafa}\cite{MV} in the light-cone gauge.
In section \ref{sec:F and IIB} we discuss some aspects of the matrix models
from the viewpoint of the F-theory interpretation of our model. Our arguments
are concerning the relation of the reduced model to the matrix model of
M-theory and the $SL(2,Z)$ symmetry of type IIB string theory.

In section \ref{sec:BP and GC} we analyze the general covariance of the matrix
model in a specific class of backgrounds. We show the physical equivalence of
the backgrounds in this class connected by the general coordinate
transformations. The topological symmetry plays a central role in this
analysis. Although our analysis is limited to the backgrounds of commuting
matrices, this supports the expectation that our model as a whole is
independent
of the backgrounds by virtue of the topological symmetry.

While carrying out the present work, we became aware of the work
\cite{HofPark}, in which they also constructed a topological matrix model
starting from the fundamental principle in a spirit quite close to ours.



\section{{\sc Topological Model}}
\label{sec:TM}

Topological model is the most promising candidate for the background
independent
description of the underlying theory of strings \cite{WittenTO}\cite{WittenTP}.
As stated in the introduction, the constituents of our model are fuzzy
instantons represented by the coordinates of a noncommutative spacetime,
$N\times N$ hermitian matrices
$X^{\mu}$, where the index $\mu$ runs from $0$ to $9$.\footnote{The dimension
and the signature of spacetime are constrained to some extent, as we will
discuss below.} We shall hypothesize that the underlying symmetry is a
topological symmetry, that is, arbitrary deformations of the noncommutative
coordinates $X^{\mu}$:
\begin{equation}
\delta X^{\mu} = {\epsilon}^{\mu}.
\label{eqn:topsym}
\end{equation}

\noindent 
where ${\epsilon}^{\mu}$'s are arbitrary $N\times N$ hermitian matrices.

In what follows we will construct a topologically twisted version of
supersymmetric reduced model with a certain modification. We shall take the
action which has the large symmetry (\ref{eqn:topsym}) to be identically zero,
\begin{equation} S = 0,
\end{equation}

\noindent and carry out the BRST gauge fixing of the symmetry
(\ref{eqn:topsym})
with this action, according to \cite{BauSin} and \cite{BMS}. Now let us
introduce
the BRST transformation laws,
\begin{eqnarray}
\delta X^{\mu} = \psi^{\mu}, \qquad \delta\psi^{\mu} = 0,
\end{eqnarray}

\noindent where the fields $X^{\mu}$ and $\psi^{\mu}$ are $N\times N$ hermitian
matrices and the ghost numbers of them are $0$ and $1$, respectively. Since the
gauge fixed action associated with this BRST symmetry will have residual gauge
symmetries, we will perform a second stage of gauge fixing
later.\footnote{Strictly speaking, they are not gauge symmetries, because the
base manifold of our model is a point and there are no local symmetries.}

To fix the topological symmetry, it seems natural to choose certain
``self-dual" equations as the gauge conditions. The ``self-dual" equation is a
variant of the higher dimensional analogue of the self-dual equation in four
dimensions given in
\cite{CDFN}:
\begin{equation} [X^{\mu},X^{\nu}] = \half
T^{\mu\nu\rho\sigma}[X_{\rho},X_{\sigma}].
\label{eqn:SDE}
\end{equation}

\noindent  Here we define the totally antisymmetric tensor
$T^{\mu\nu\rho\sigma}$ as
\begin{equation} T^{\mu\nu\rho\sigma} = (\zeta^T,0)
\Gamma^{\mu\nu\rho\sigma}\left(
\begin{array}{c}
\zeta \\ 0
\end{array}
\right),
\label{eqn:4AS}
\end{equation}

\noindent
where $\Gamma^{\mu\nu\rho\sigma}$ is the totally antisymmetric product
of $\Gamma$ matrices for $SO(9,1)$ spinor representation, and $\zeta$ is a unit
constant Majorana-Weyl spinor, $\zeta^T \zeta =1$. We will further impose that
$\zeta$ satisfies a $SO(8)$ Weyl condition.

\noindent
Let us decompose the gamma matrices $\Gamma^{\mu}$, in terms of the $SO(8)$
ones $\gamma^i$, into $\Gamma^0 = i\sigma_2 \otimes\mbox{\bf 1}_{16}$,
$\Gamma^i =\sigma_1 \otimes\gamma^i$, and $\Gamma^9 = \sigma_1
\otimes\gamma^9$. Then
one can easily find that the $4$-th rank antisymmetric tensor
(\ref{eqn:4AS}) is broken up into
\begin{eqnarray} T^{0ijk} = T^{0ij9} = T^{ijk9} = 0,
\qquad T^{ijkl} = \zeta^T \gamma^{ijkl} \zeta,
\label{eqn:so8decomp}
\end{eqnarray}

\noindent
where $i,j,k,l = 1,\cdots ,8$ and the tensor $T^{ijkl}$ is invariant
under $SO(7)$ rotation as is obvious from its definition. The appearance of
$SO(7)$ is understood from the construction of the ``self-dual" equation in
eight dimensions discussed in detail in \cite{CDFN}.

\noindent As the result we obtain the following explicit expression for our
gauge
conditions (\ref{eqn:SDE}):
\begin{equation}
\begin{array}{c} F_{09} = F_{0i} = F_{9i} = 0, \\
\end{array}
\label{eqn:lcctr}
\end{equation}
\begin{equation}
\left\{
\begin{array}{c}
F_{12} + F_{34} + F_{56} + F_{78} = 0, \nn\\
F_{13} + F_{42} + F_{57} + F_{86} = 0, \nn\\
F_{14} + F_{23} + F_{76} + F_{85} = 0, \nn\\
F_{15} + F_{62} + F_{73} + F_{48} = 0, \nn\\
F_{16} + F_{25} + F_{38} + F_{47} = 0, \nn\\
F_{17} + F_{82} + F_{35} + F_{64} = 0,
\label{eqn:expSDE} \\
F_{18} + F_{27} + F_{63} + F_{54} = 0, \nn
\end{array}
\right.
\end{equation}

\noindent where we define the field strengths $F_{\mu\nu} =
i[X_{\mu},X_{\nu}]$.

Some remarks are in order: (i) The latter set of the gauge conditions
(\ref{eqn:expSDE}) is used in the context of the cohomological Yang-Mills
theory
in eight dimensions \cite{BKS}, where they constructed the nearly topological
Yang-Mills theory, in particular, on the Joyce manifold with
$spin(7)$ holonomy.\footnote{See also \cite{AO} for related discussions for the
``self-dual" equations.} (ii) The second rank tensor $F_{ij}$ in eight dimensions
belongs to {\bf 28} of $SO(8)$, whose $SO(7)$ decomposition is {\bf 21}
$\oplus$
{\bf 7}. A set of seven equations (\ref{eqn:expSDE}) belongs to the {\bf 7},
and
remarkably it enjoys the octonionic structure as noted in \cite{CDFN}. The
appearance of the octonion may explain that the dimensions $9+1$ of spacetime
is
critical, including the signature, along the line of the argument given in
\cite{KugoTown}.\footnote{As advertised in the abstract, the extra $1+1$
dimensions arise automatically in our formulation, in addition to 9+1
dimensions. Thus this may in turn account for that the critical dimension is
$10+2$.} (iii) Since the base manifold of our model is a point, the
``self-dual"
equation does not necessarily mean the instanton equation in the field theory
sense, rather it is only formal analogue of that in higher dimensions. We note,
however, that there are possibilities the ``self-dual" equation does indeed
become the instanton equation in the field theory sense, if we take certain
large N limits which give, say, the configurations of matrices $X$'s
corresponding to the toroidal compactifications of the matrix model of
M-theory.\cite{BFSS}\cite{WT4}

In order to construct a gauge fixed action for the topological symmetry, we
must
introduce the antighosts $\chi_{\mu\nu}$ with the ghost number $-1$ and the
Nakanishi-Lautrup fields $b_{\mu\nu}$, whose BRST transformation rules are
\begin{equation}
\delta\chi_{\mu\nu} = ib_{\mu\nu}, \qquad \delta b_{\mu\nu} = 0.
\label{eqn:topantiaux}
\end{equation}

\noindent
They satisfy the ``anti self-dual" equations:
\begin{equation}
\chi_{ij} = -{1 \over 6}T_{ijkl}\chi^{kl}, \qquad
b_{ij} = -{1 \over 6}T_{ijkl}b^{kl}, \label{eqn:chi-b}
\end{equation}

\noindent
or equivalently,
\begin{eqnarray}
{1\over 4}\left(\delta_{ik}\delta_{jl} - {1\over 2}T_{ijkl}\right)\chi^{kl}
&=& \chi_{ij},\\
{1\over 4}\left(\delta_{ik}\delta_{jl} - {1\over 2}T_{ijkl}\right)b^{kl}
&=& b_{ij}.
\end{eqnarray}

\noindent
The operator $P_{ijkl}={1\over 4}\left(\delta_{ik}\delta_{jl} - {1\over
2}T_{ijkl}\right)$ is a projection operator onto the subspace of the eigenvalue
$-3$ of $T_{ijkl}$.\footnote{The projection operator for the eigenvalue $1$ is
${3\over 4}\left(\delta_{ik}\delta_{jl} +
{1\over 6}T_{ijkl}\right)$. In order to write the projection operators in
$SO(9,1)$ covariant way, we need to make them of a quadratic form in
$T^{\mu\nu\rho\sigma}$, because there are three distinct eigenvalues of
$T^{\mu\nu\rho\sigma}$.}

\noindent
Now the gauge fixed action is
\begin{eqnarray}
S_{GF} &=& -i
\delta\Tr\left\{ {1 \over 4}\chi_{\mu\nu}\left(F^{\mu\nu} - \half
T^{\mu\nu\rho\sigma}
F_{\rho\sigma}\right) + \half\alpha_{\mu\nu}\chi_{\mu\nu}b^{\mu\nu}
\right\},
\label{eqn:GFA}
\end{eqnarray}

\noindent where $\alpha_{\mu\nu}$'s are gauge fixing parameters and they are
not components of a second rank tensor. The normalization of the action is in
conformity with that of the projection operator $P_{ijkl}$.

\noindent
For later purpose we take the Landau gauge for the gauge fixing
functions of (\ref{eqn:lcctr}), $\alpha_{09} = \alpha_{0i} = \alpha_{9i} = 0$,
and set the gauge parameters $\alpha_{ij} = \alpha$ for those of
(\ref{eqn:expSDE}). Then integrating out the auxiliary fields $b_{09}$,
$b_{0i}$, and $b_{9i}$, the gauge fixed action
(\ref{eqn:GFA}) reduces to
\begin{eqnarray}  S_{GF} &=& \Tr\biggl\{{1 \over 4}b_{ij}
\left(F^{ij} - \half T^{ijkl}F_{kl}\right) +{\alpha \over 2}b_{ij}b^{ij}
 -{1 \over 4}\chi_{ij}\left( [X^{[i},\psi^{j]}] - \half
T^{ijkl}[X_{[k},\psi_{l]}]
\right) \nn\\
 & &  - 2\chi_{09}[X^{[0},\psi^{9]}] - 2\chi_{0i}[X^{[0},\psi^{i]}] -
2\chi_{9i}[X^{[9},\psi^{i]}]\biggr\}\\
&=&\Tr\biggl\{b_{ij}F^{ij} + {\alpha \over 2}b_{ij}P^{ijkl}b_{kl}
-\chi_{ij}[X^{[i},\psi^{j]}] \nn\\
& &- 2\chi_{09}[X^{[0},\psi^{9]}] - 2\chi_{0i}[X^{[0},\psi^{i]}] -
2\chi_{9i}[X^{[9},\psi^{i]}]\biggr\},\label{eqn:GFArdc}\\
&\mbox{with}&\!\!\!\mbox{the constraints}\qquad
\delta([X^0,X^9])\times\delta([X^0,X^i])\times\delta([X^9,X^i]).\nn
\end{eqnarray}

\noindent Now let us look on the delta function constraints,
\begin{equation}  [X^0,X^9] = 0, \quad [X^0,X^i] = 0, \quad [X^9,X^i] = 0.
\label{eqn:delta}
\end{equation}

\noindent Using a $U(N)$ gauge rotation, we can choose a basis in which $X^0$
and $X^9$  take their values on the Cartan subalgebra of $U(N)$. In such a
basis
the second and third equations in (\ref{eqn:delta}) constrain the values of
$X^i$'s on the Lie algebra of $U(N-M)\otimes U(1)^M , (M = 0, \cdots ,N)$, depending on the values of $X^0$ and $X^9$. For later
convenience, we will denote the gauge groups $U(N-M)\otimes U(1)^M , (M = 0, \cdots ,N)$ as ${\cal U}(N)$ collectively. Having this in
mind and taking into account the fact that the integrations over the fermions
in the second line of the gauge fixed action (\ref{eqn:GFArdc}) give the
Jacobian
factors for the delta functions, we can further reduce the gauge fixed action
to
\begin{equation} S_{GF} = \Tr\biggl\{b_{ij}
F^{ij} +{\alpha \over 2}b_{ij}P^{ijkl}b_{kl}
-\chi_{ij}[X^{[i},\psi^{j]}]\biggr\},
\label{eqn:GFaction}
\end{equation}

\noindent where all fields take their values on the Lie algebra of
${\cal U}(N)$.

Note that $X^0$ and $X^9$ do not completely disappear from the system. The
diagonal part of them do remain, and in some cases left their traces of $U(1)$
factors in
the gauge symmetry. This is reminiscent of the light-cone gauge of string
theory, in which only do the zero modes of the light-cone coordinates survive.
Indeed it will be suggested that we can regard this gauge as the
light-cone gauge of F-theory, as we will discuss in section \ref{sec:F and
IIB}.

As mentioned previously, the gauge fixed action (\ref{eqn:GFaction}) for the
topological symmetry  has a fermionic gauge symmetry,
$\delta_{\lambda}\psi^i = [X^i, \lambda], \delta_{\lambda}b_{ij} =
\{\chi_{ij},\lambda\}$, and $\delta_{\lambda}X^i =
\delta_{\lambda}\chi_{ij} = 0$, where $\lambda$ is a Grassmann-valued matrix.
In fact the variation of the action under this transformation is BRST-exact. In
order to fix this symmetry, we must introduce a ghost $\phi$ for ghost
$\psi^i$,
whose BRST transformation laws are
\begin{eqnarray}
\tilde{\delta}X^i &=& 0, \qquad \tilde{\delta}\psi^i = [X^i,\phi],
\qquad \tilde{\delta}\phi = 0, \nn\\
\tilde{\delta}\chi^{ij} &=& 0,
\qquad \tilde{\delta}b_{ij} = -i[\chi_{ij},\phi]
\qquad (\delta\phi = 0).
\label{eqn:BRSTsecond}
\end{eqnarray}

\noindent We will choose the following gauge function for this symmetry,
\begin{equation} [X_i,\psi^i],
\label{eqn:psiGFF}
\end{equation}

\noindent and introduce an antighost ${\bar \phi}$ and an auxiliary field
$\eta$
with the BRST transformation rules,
\begin{equation}
\tilde{\delta}{\bar \phi} = 2\eta,
\qquad \tilde{\delta}\eta = \half[\bar{\phi},\phi],
\qquad (\delta{\bar \phi} = \delta\eta = 0).
\label{eqn:BRSTsecondtwo}
\end{equation}

\noindent Then the total gauge fixed action is
\begin{eqnarray}
\tilde{S}_{GF} &=& S_{GF} - (\delta + \tilde{\delta})
\Tr\left(\half\bar{\phi}[X_i,\psi^i] - {1 \over 4}\bar{\phi}[\phi,\eta]
-{i \over 4}b_{ij}\chi^{ij}\right) \nn\\
&=& \Tr\biggl\{b_{ij}F^{ij} +{\alpha \over 2}b_{ij}P^{ijkl}b_{kl}
-\chi_{ij}[X^{[i},\psi^{j]}] \nn\\
& & - \eta[X_i,\psi^i] -\half \bar{\phi}\{\psi_i,\psi^i\}
-\half\phi\{\eta,\eta\} -{1 \over 4}\phi\{\chi_{ij},\chi^{ij}\}
\label{eqn:totGFAproto}\\
& & +\half[X_i,\phi][X^i,\bar{\phi}] +{1 \over 8}[\phi,\bar{\phi}]^2
-{1 \over 4} b_{ij}P^{ijkl}b_{kl}
\biggr\} \nn,
\end{eqnarray}

\noindent
where the total BRST operator $\delta + \tilde{\delta}$ is nilpotent
up to a ${\cal U}(N)$ gauge transformation. We have added suitable BRST exact
terms in order to make the action of the standard form.

At this stage we set the gauge parameters $\alpha$ to be zero in such a way
that the coefficient of the term $b_{ij}P^{ijkl}b_{kl}$ becomes $-1/4$. Thus
our choice of gauge for the gauge conditions (\ref{eqn:expSDE}) is the Feynmann
gauge. Then integrating out the auxiliary fields $b_{ij}$, the total gauge
fixed action reduces to
\begin{eqnarray}
\tilde{S}_{GF} &=& \Tr\biggl\{{1 \over 4}F_{ij}F^{ij} -
{1 \over 8}T^{ijkl}F_{ij}F_{kl}
-\chi_{ij}[X^{[i},\psi^{j]}]
\nn\\
& & - \eta[X_i,\psi^i] -\half\bar{\phi}\{\psi_i,\psi^i\}
-\half\phi\{\eta,\eta\} -{1 \over 4}\phi\{\chi_{ij},\chi^{ij}\}
\label{eqn:totGFA}\\
& & +\half[X_i,\phi][X^i,\bar{\phi}] +{1 \over 8}[\phi,\bar{\phi}]^2
\biggr\} \nn.
\end{eqnarray}

\noindent
This gauge fixed action still has the ordinary gauge symmetry. The gauge fixing
procedure is quite standard and we will not carry out it here. We would,
however, like to note that we can make the total BRST operator nilpotent off
shell by
adding the BRST operator for the ordinary gauge symmetry.

\parbigskipn
\underline{{\it The Relation to The Supersymmetric Reduced Model}}
\parmedskipn

Now let us discuss the relation of our model to the supersymmetric reduced
model. The fields in the latter are bosons which transform as a ({\bf 9},{\bf
1}) vector under a global $SO(9,1)$ rotation, and fermions as a {\bf 16}
spinor.
Under a subgroup $SO(1,1)\otimes SO(7)$ of $SO(9,1)$, they are decomposed into
\begin{eqnarray}
(\mbox{\bf 9},\mbox{\bf 1}) &\longrightarrow&
\mbox{\bf 8}_{\mbox{\bf 0}} \oplus \mbox{\bf 1}_{\mbox{\bf 2}}
\oplus \mbox{\bf 1}_{\mbox{\bf -2}}, \\
\mbox{\bf 16} &\longrightarrow&
\mbox{\bf 8}_{\mbox{\bf 1}} \oplus \mbox{\bf 7}_{\mbox{\bf -1}}
\oplus \mbox{\bf 1}_{\mbox{\bf -1}},
\end{eqnarray}

\noindent where the subscripts denote twice $SO(1,1)$ spin, and we embedded
$SO(7)$ into one of the spinor representations of $SO(8)$, say, ${\bf 8}_{{\bf
s}}$.

On the other hand the fields in the former are the following:
\begin{eqnarray}
&&X^i \,\, (\mbox{\bf 8}_{\mbox{\bf 0}}), \qquad
\phi\,\,(\mbox{\bf 1}_{\mbox{\bf 2}}), \qquad
\bar{\phi}\,\,(\mbox{\bf 1}_{\mbox{\bf -2}}), \\ &&\psi^i\,\,(\mbox{\bf
8}_{\mbox{\bf 1}}),\qquad
\chi_{ij}\,\,(\mbox{\bf 7}_{\mbox{\bf -1}}), \qquad
\eta\,\,(\mbox{\bf 1}_{\mbox{\bf -1}}),
\end{eqnarray}

\noindent where the subscripts denote the ghost number.

Thus we find that the field contents of our model completely match with those
of the reduced model. This shows that we can identify our model as a
topologically twisted version of the supersymmetric reduced model. Indeed one
can see that the action (\ref{eqn:totGFA}) is equivalent to that of the
supersymmetric reduced model
\begin{equation}
S_{RM} = \Tr\biggl\{-{1 \over 4}[A_{\mu},A_{\nu}]^2
-\half\bar{\Psi}\Gamma^{\mu}[A_{\mu},\Psi]\biggr\},
\label{RM}
\end{equation}

\noindent
up to the term $\Tr T^{ijkl}F_{ij}F_{kl}$, by the following identifications of
the fields:
\begin{eqnarray}
A^i &=& X^i,\qquad A_0 + A_9 = \phi,\qquad A_0 - A_9 =\bar{\phi}\nn\\
\lambda_{+}^i &=& \psi^i,\qquad \lambda_{-}^a = 2\chi^{8a},
\qquad \lambda_{-}^8 = \eta,
\label{eqn:fieldid}
\end{eqnarray}

\noindent
where $a = 1,\cdots,7$ and the $SO(8)$ chiral spinors $\lambda_{+}$ and
$\lambda_{-}$ are given by
\begin{equation}
\Psi^T = (\lambda_{+}^T,\lambda_{-}^T,\mbox{\bf 0},\mbox{\bf 0}),
\label{eqn:lambda}
\end{equation}

\noindent
in the convention of the gamma matrices $\Gamma^{\mu}$ employed
before.\footnote{$\Gamma^0 = i\sigma_2 \otimes\mbox{\bf 1}_{16}$, $\Gamma^i
=\sigma_1 \otimes\gamma^i$, and $\Gamma^9 = \sigma_1 \otimes\gamma^9$.}

\noindent
Under the identification (\ref{eqn:fieldid}) of the fields, the resulting
action is written as
\begin{eqnarray}
S_{RM} &=& \Tr\biggl\{{1 \over 4}F_{ij}F^{ij}
-\chi_{8a}\left(2[X^{[8},\psi^{a]}] +c^{abc}[X_{[b},\psi_{c]}]
\right)- \eta[X_i,\psi^i]\nn\\
& &  -\half\bar{\phi}\{\psi_i,\psi^i\}
-\half\phi\{\eta,\eta\} -2\phi\{\chi_{8a},\chi^{8a}\}
\label{eqn:Actionocto}\\
& & +\half[X_i,\phi][X^i,\bar{\phi}] +{1 \over 8}[\phi,\bar{\phi}]^2
\biggr\} \nn.
\end{eqnarray}

\noindent
Here we have used specific representations of the gamma matrices $\gamma^i$ and
of the auxiliary fields $\chi_{ij}$ in terms of the structure constants
$c_{abc}$ for octonions. They are summarized in the appendix.

Note that the signature of $SO(1,1)$ is relevant since the generator of
$SO(1,1)$ is nothing other than the ghost number current and thus must
correspond to a scale, not a phase, transformation. This shows the signature
of ($\phi + \bar{\phi},\phi - \bar{\phi}$) must be ($1,1$).

We would also like to remark that the total gauge fixed action
(\ref{eqn:totGFA}) contains an extra term $\Tr T^{ijkl}F_{ij}F_{kl}$, compared
with the supersymmetric reduced model. For finite $N$ this term vanishes by
virtue of the cyclicity of the trace and the Jacobi identity. It can, however,
survive in certain large $N$ limits. For instance the configurations of
matrices
corresponding to the toroidal compactifications break the cyclicity of the
trace. Moreover, in this case, the Jacobi identity can be lifted to the Bianchi
identity, and it is well-known that the Bianchi identity does not hold in the
presence of the topological defects. Thus our model is endowed with an
interesting modification of the supersymmetric reduced model in the large $N$
limits.

One may, however, suspect that this extra term would violate the successful
outcomes of the matrix models, such as the emergence of various brane solutions
and the precise agreements of the brane-brane scatterings in the matrix models
with those in supergravities. Fortunately this does not seem to happen. This is
because the variations of the extra term with respect to $X$'s are vanishing as
far as the Bianchi identity holds.
Therefore, in the case of simple brane configurations such as $[P,Q]=
\mbox{const.}$ and its generalizations, the equations of motion of the matrix
models are not altered, and the extra term is independent of the detailed form
of the matrices. Thus it appears that this term would not affect the successful
results for the known brane solutions and the amplitudes of the brane-brane
scatterings.

We would, however, like to emphasize that there could be some effects of the
extra term $\Tr T^{ijkl}F_{ij}F_{kl}$ on the dynamics of the matrix model, if
the configurations of matrices break the Bianchi identity. In this case even
the equations of motion are modified and we expect that the extra term would
produce brane solutions yet unknown or missing.



\section{{\sc F-theory Interpretation and IIB String}}
\label{sec:F and IIB}

In addition to the ($9+1$)-dimensional coordinates $X^{\mu}$, there emerge two
extra bosons $\phi$ and $\bar{\phi}$ with the signature ($1,1$). This is likely
to indicate that the spacetime dimensions of our model are promoted to $10+2$.
Thus we are tempted to interpret our model as a possible description of
F-theory
\cite{Vafa}\cite{MV}. As we took the light-cone gauge in our formulation, we
suggest that our model would describe the light-cone F-theory with $9+1$
transverse dimensions. This viewpoint may illuminate the understanding of some
aspects of the matrix models.

\parbigskipn
1. {\it The Relation of The Reduced Model to The Matrix Model of
M-Theory}
\parmedskip

So far there are no arguments to directly connect F-theory with M-theory
without
compactifying M-theory. Our viewpoint, however, may provide a way to relate
them
directly. One naively expect that the compactification of one time direction of
F-theory leads to M-theory in $10+1$ dimensions. Indeed our viewpoint supports
this idea as follows:

Let us look on the bosons in our model. We will list them below.
\begin{equation}
\begin{array}{cc|cc|}
(X^0,X^9) &\qquad\qquad & X^i & (\phi +\bar{\phi},\phi
-\bar{\phi}) \nn\\ 1+1 &\qquad\qquad & 8 & 1+1 \nn\\
\mbox{light-cone} &\qquad\qquad & \mbox{transverse} & \mbox{transverse} \\
\end{array}
\end{equation}

\noindent The reduced model is lifted to a ($1+0$)-dimensional supersymmetric
Yang-Mills theory ($SYM_{1+0}$) by the compactification of one time direction
$\phi +\bar{\phi}$, in which the matrix $\phi +\bar{\phi}$ is represented by a
covariant derivative $-i\del_{t} -A_0 (t)$ \cite{BFSS}\cite{WT4}. Since
$SYM_{1+0}$ is nothing other than the matrix model of M-theory with $9+0$
transverse directions, this shows the relation of M-theory to F-theory
compactified on a timelike
$S^1$:
\begin{equation}
\mbox{LC F} \stackrel{S^1}{\longrightarrow}\mbox{LC M},
\end{equation}

\noindent
where LC denotes the light-cone gauge and $S^1$ is in a timelike
direction.

\noindent
The connection of the reduced model with the matrix model of M-theory
was also anticipated in \cite{IKKT1}.

\parbigskipn
2. {\it SL(2,Z) Symmetry of Type IIB String}
\parmedskip
Type IIB string theory would be obtained by the compactification of
F-theory on a ($1,1$) space \cite{Vafa}.\footnote{The authors in \cite{LMS}
made a quite different proposal in this respect.} In the light of our F-theory
interpretation, a
matrix description of the light-cone type IIB theory is expected to be the
reduced model on $T^{1,1}$ torus, where $\phi$- and $\bar{\phi}$-directions are
compactified. Thus the light-cone type IIB theory seems to be described by a
($1+1$)-dimensional
$N=8$ supersymmetric Yang-Mills theory ($SYM_{1+1}$) discussed as the type IIA
string in \cite{Motl}\nocite{DVV}--\cite{BanksSeib}. In the context of the
matrix model of M-theory, the authors in \cite{SethiSuss}\cite{BanksSeib}
discussed that the type IIB string would be
reproduced by a ($2+1$)-dimensional supersymmetric Yang-Mills theory on a
$2$-torus, in which one cycle of the torus was taken much smaller than the
other. Although our viewpoint is different from theirs, we expect that both of
them would be linked to each other.

To precisely interpret $SYM_{1+1}$ as the light-cone type IIB string (both D-
and F-), we have to explain how the correct chirality comes about in a
$T^{1,1}$ compactification. Here we only assume that the correct chirality
could be obtained in our framework.

Now let us discuss the $SL(2,Z)$ symmetry of type IIB strings. As conjectured
in \cite{Vafa}, the $SL(2,Z)$ symmetry is expected to be understood as a
geometrical symmetry of the torus $T^{1,1}$. In the case at hand we might be
able to support this conjecture in the following way:

Let the radius of ($\phi +\bar{\phi},\phi -\bar{\phi}$) be ($R_{+},R_{-}$). The
moduli parameter of the torus $T^{1,1}$ is given by $\tau = iR_{+}/R_{-}$. The
compactification of the reduced model on
$S^1$ with the radius $R_{+}$ leads to the matrix model of M-theory as
mentioned above. Subsequently we compactify the matrix model of M-theory on
$S^1$ with the radius $R_{-}$. Then the radius $R_{-}$ is related to the string
coupling constant $g_s$ via $R_{-} =g_s$ in string unit. Thus a modular
transformation $\tau\to -1/\tau$ gives the S-duality $g_s \to 1/g_s$ as we
expected.

\parbigskipn
A remark is in order. Type IIB superstring field theory was derived
from the IIB matrix model in \cite{IKKT2}.\footnote{We would like to thank
Asato Tsuchiya for explaining their work.} They performed the light-cone
decomposition of the coordinates in order to connect the IIB matrix model with
the light-cone string field theory. They, however, did not take the light-cone
gauge, rather they made only the formal light-cone decomposition of the
coordinates. The degrees of freedom of the light-cone directions were not
subtracted at all.  This suggests that the reduced model with
($9+1$)-dimensional spacetime contains only the degrees of freedom of the light-cone formulation.\footnote{This point was also noted in \cite{Hama} in the light of supersymmetry transformation of the Wilson loops.}
Thus our F-theory interpretation is not led to an immediate contradiction to
the result obtained in \cite{IKKT2}.



\section{{\sc Broken Phase and General Covariance}}
\label{sec:BP and GC}

In this section we would like to argue how we should understand the general
covariance of the matrix model in our framework. Our analysis here is limited
to
that in a broken phase, that is, in a specific background configuration of the
matrices. The understanding of the broken phase itself is important to connect
our model with the physics in the real world. We, however, consider it to be
complementary to the understanding of the background independence of the matrix
model as well.

Now let us take a background in which the matrices are mutually commuting:
\begin{equation} [x^i,x^j] = [x^i,\varphi] = [x^i,\bar{\varphi}]  =
[\varphi,\bar{\varphi}] = 0,
\label{eqn:bgd}
\end{equation}

\noindent where $x^i$, $\varphi$, and $\bar{\varphi}$ are the background fields
for $X^i$,
$\phi$, and $\bar{\phi}$ respectively.

\noindent This background is considered as a commutative spacetime limit of a
noncommutative one. Thus we should be able to see the general covariance of the
ordinary spacetime.

As discussed in section \ref{sec:TM}, the symmetries of our model are a
topological symmetry ${\cal T}$ and the gauge symmetries ${\cal G}$. The
background we took breaks the symmetries down to a gauge symmetry
${\cal H}$, which commutes with the background. Under the transformations
\begin{equation}
\delta X^i = f^i(x^i,\varphi,\bar{\varphi}), \qquad
\delta \bar{\phi} = f_{\bar{\varphi}}(x^i,\varphi,\bar{\varphi}),
\label{eqn:TGCT}
\end{equation}

\noindent
for the fluctuations from the background, where $f^i$ and $f_{\bar{\varphi}}$
are arbitrary functions of the background matrices
$x^i$, $\varphi$, and $\bar{\varphi}$, the gauge symmetry ${\cal H}$ is
preserved. Note that a ghost of ghost $\phi$ is invariant under the topological
transformation. Now these transformations induce replacements of the
background
\begin{equation}
x^i \to x^i + f^i(x^i,\varphi,\bar{\varphi}), \qquad
\bar{\varphi} \to \bar{\varphi} + f_{\bar{\varphi}}(x^i,\varphi,\bar{\varphi}),
\label{eqn:GCT}
\end{equation}

\noindent
which are nothing other than the general coordinate transformations. The above
transformations (\ref{eqn:TGCT}) are caused by the generators of a broken
symmetry, the topological symmetry
${\cal T}$. This means that the backgrounds connected with the transformations
(\ref{eqn:GCT}) are equivalent to each other up to the BRST transformations.
Thus they all are physically equivalent. This shows the general covariance of
the matrix model in a broken phase. The key ingredient here is the topological
symmetry ${\cal T}$.

We should not, however, expect the full general covariance of the
($10+2$)-dimensional spacetime because of the invariance of $\phi$ under the
topological transformation. We also remark that our derivation has been
performed in the light-cone gauge. However the inclusion of the light-cone
coordinates seems to have no obstruction for the above arguments. We do not now
have a definite answer to a question whether the general covariance we
discussed is simply ($9+1$)- or intricately ($10+1$)-dimensional one. We would
also like to mention that the analysis in this
section indicates the topological symmetry ${\cal T}$ is a key to
understand the background independence of the matrix model, as it should be.



\section{{\sc Conclusions and Discussions}}
\label{sec:CD}

We proposed a candidate for the primal principle to define the underlying
theory
of string theory. Our first hypothesis is that the most fundamental
constituents
of the theory are fuzzy instantons represented by the matrices which denote a
position in a noncommutative spacetime. The second one is that the theory has a
topological symmetry, that is, arbitrary deformations of the fields, which is
likely to be the maximum one considered naturally. The resulting theory is a
topologically twisted version of the supersymmetric reduced model with a
certain
modification. There emerged extra $1+1$ spacetime dimensions in addition to the
starting $9+1$ dimensions, thereby we were tempted to interpret our model as a
matrix description of F-theory.

Until now F-theory is defined only through the compactifications on
elliptically
fibered complex manifolds. To verify our interpretation, we must study the
compactifications of our model on elliptically fibered surfaces. In this
connection, it would be interesting to study the compactification of our model
on a $K_3$ orbifold, $T^4/Z_2$ \cite{Sen}.

Although we took the light-cone gauge in our formulation, it does not seem to
be
difficult to construct our model in covariant way. This appears  to be easily
realized by taking the Feynmann gauge for all the gauge conditions, instead of
taking the Landau gauge for those containing the light-cone coordinates. This
is
only a choice of gauge, and thus two formulations would be physically
equivalent. We expect that a covariant formulation of our model sheds some
lights on that of the matrix model of M-theory.

Far from being a problem, there is another formulation to obtain a
topologically
twisted supersymmetric reduced model. Starting with only transverse
$8$-dimensional space, we can construct almost the same model as the one in the
light-cone gauge. The difference arises in the gauge group. This alternative
formulation gives the model with $U(N)$ gauge symmetry, not
${\cal U}(N)$. The formulation itself is mathematically more beautiful
than that performed in the present paper. In this formulation, the
($9+1$)-dimensional whole world is emerged as a hologram of the $8$-dimensional
transverse space. This gives a concrete realization of the world as a hologram
\cite{tHooft}\cite{Susskind}. It is amusing but quite different from the one in
the present work to interpret our model in this manner.

We would like to mention the physical states of our model. They are defined as
the states that are invariant under the BRST transformation up to a gauge
transformation. Since the BRST charge is a supercharge which is singlet under
$SO(7)$ rotation, the physical states are all BPS states that preserves at
least
$1/16$ of the supersymmetry. There are no non-BPS states in our model.
Considering the connection to the real world, the BRST symmetry must be broken
spontaneously. By this mechanism four dimensional spacetime should be
dynamically generated. This problem extremely deserves to be investigated
further.

We discussed the general covariance of the matrix model. Although our analysis
was limited to that in a broken phase, we want to study the general covariance
in the unbroken phase. The topological symmetry must be a key to show the
background independence of the matrix model. Now we believe that the background
independence has been already encoded in our formulation of the matrix model.



\section*{{\sc Acknowledgements}}

We would like to thank Yoichi Kazama and Tamiaki Yoneya for stimulating
discussions on various subjects in string, D-brane and matrix model.
S.H.\ would like to thank Kazutoshi Ohta and Akira Tokura for valuable
discussions. He is also grateful to Hiroshi Itoyama and Keiji Kikkawa for
encouragement.
M.K.\ is obliged to Yuji Okawa for sharing insight on non-commutative space
and Miao Li for comments.
The work of S.H.\ was supported in part by Osaka University Yukawa Memorial
Foundation and by Soryushi Shogakukai.
That of M.K.\ was supported in part by the Grant-in-Aid for Scientific
Research by the Ministry of Education, Science, Sports and Culture
(\#09640337).



\section*{{\sc Appendix}}
\setcounter{equation}{0}
\renewcommand{\theequation}{A.\arabic{equation}}

In this appendix, we will make a brief summary of some definitions and formulae
concerning the octonion used in section \ref{sec:TM}.

The octonion basis, $e_a \, (a = 1,\cdots,7)$ and $e_8 = 1$, satisfy
\begin{equation}
e_a e_b = -\delta_{ab} + c_{abc}e_c,
\label{eqn:octonion}
\end{equation}

\noindent
where $c_{abc}$ are the structure constants for octonions and totally
antisymmetric.

\noindent
Next we will define the following $8\times 8$ matrices:
\begin{eqnarray}
(t^a)_{bc} &=& c_{abc}, \nn\\
(t^a)_{b8} &=& -(t^a)_{8b} = \delta_{ab},
\label{eqn:tmatrix}\\
(t^a)_{88} &=& 0, \nn\end{eqnarray}

\noindent
The $SO(8)$ gamma matrices $\gamma^i$ are expressed, in terms of these $8\times
8$ matrices, as
\begin{eqnarray}
\gamma^a =
\left(
\begin{array}{cc}
\mbox{\bf 0} & t^a \\
-t^a & \mbox{\bf 0}
\end{array}
\right),
\qquad
\gamma^8 =
\left(
\begin{array}{cc}
\mbox{\bf 0} & \mbox{\bf 1}_8 \\
\mbox{\bf 1}_8 & \mbox{\bf 0}
\end{array}
\right).
\label{eqn:gammatexp}
\end{eqnarray}

\noindent
Then one can find that the $4$-th rank antisymmetric tensors $T^{ijkl}$ defined
in (\ref{eqn:so8decomp}) enjoy an expression in terms of the structure
constants $c_{abc}$, when picking a Majorana-Weyl spinor $\zeta_{\alpha} =
\delta_{8\alpha}$:
\begin{eqnarray}
T^{8abc} &=& c_{abc} \nn\\
T^{abcd} &=& {1 \over 3}\left(-c_{abe} c_{ecd} + c_{ace} c_{ebd}
-c_{ade} c_{ebc}\right).
\label{eqn:Texpc}
\end{eqnarray}

\noindent
Note that in this representation of $T^{ijkl}$ the gauge conditions
(\ref{eqn:expSDE}) are given succinctly by
\begin{equation}
F_{8a} = \half c_{abc} F^{bc},
\label{eqn:gaugecondocto}
\end{equation}

\noindent
and the auxiliary fields $\chi_{ij}$ satisfy
\begin{equation}
\chi_{ab} = c_{abc}\chi^{c8}.
\label{eqn:chiocto}
\end{equation}

\noindent
Lastly we will list two useful identities below:
\begin{eqnarray}
c_{acd} c_{bcd} &=& 6\delta_{ab}\quad
\Leftrightarrow \quad
\mbox{tr}(\{t^a,t^b\})= -16\delta_{ab},\nn\\
c_{adf} c_{bfe} c_{ced}&=& -3c_{abc}\quad
\Leftrightarrow\quad
\mbox{tr}(t^a t^b t^c ) =0.
\end{eqnarray}




\end{document}